\documentclass[aps,preprint,showpacs,showkeys]{revtex4}
\usepackage{amsmath}
\usepackage{amssymb}

\begin{document}

\title{Solution of Fokker-Planck equation for a broad class of drift and
diffusion coefficients}
\author{Kwok Sau Fa}
\affiliation{Departamento de F\'{\i}sica, Universidade Estadual de Maring\'{a}, Av.
Colombo 5790, 87020-900, \ Maring\'{a}-PR, Brazil, Tel.: 55 44 32614330,
Fax: 55 44 32634623}
\email{kwok@dfi.uem.br}

\begin{abstract}
We consider a Langevin equation with variable drift and diffusion
coefficients separable in time and space and its corresponding Fokker-Planck
equation in the Stratonovich approach. From this Fokker-Planck equation we
obtain a class of exact solutions with the same spatial drift and diffusion
coefficients. Furthermore, we analyze some details of this system by using
the spatial diffusion coefficient $D(x)=\sqrt{D}\left\vert x\right\vert ^{-%
\frac{\theta }{2}}$.
\end{abstract}

\maketitle

\newpage

In these last decades, anomalous diffusion properties have been extensively
investigated by several approaches in order to model different kinds of
probability distributions such as long-range spatial or temporal
correlations \cite{klafter,marconi}. These approaches have been used to
describe numerous systems in several contexts such as physics, hydrology,
chemistry and biology. The diffusion process is classified according to the
mean square displacement (MSD) 
\begin{equation}
\left\langle x^{2}(t)\right\rangle \sim t^{\alpha }\text{ .}  \label{eq0}
\end{equation}%
In the case of normal diffusion, the MSD grows linearly with time ($\alpha
=1 $). For $0<\alpha <1$ the process is called subdiffusive, and for $\alpha
>1$ the process is called superdiffusive. The well-established property of
the normal diffusion described by the Gaussian distribution can be obtained
by the usual Fokker-Planck equation with a constant diffusion coefficient
(without the drift term) \cite{risken,gardiner} or by an
integro-differential diffusion equation with the exponential function for
the waiting time probability distribution \cite{kwokwang}. Anomalous
diffusion regimes can also be obtained by the usual Fokker-Planck equation,
however, they arise from variable diffusion coefficient which depends on
time and/or space. On the other hand, in the view of Langevin approach it is
associated with a multiplicative noise term. In other approaches such as the
generalized Fokker-Planck equation (nonlinear) and fractional equations,
they can describe anomalous diffusion regimes with a constant diffusion
coefficient.

The Langevin equation is a very important tool for describing systems out of
equilibrium \cite{risken,gardiner}. Moreover, this equation has been
extensively investigated; many properties and analytical solutions of it
have also been revealed. In this work, we present solutions of a class of
the Langevin equation with the deterministic drift and multiplicative noise
terms in time and space. To do so, we obtain the corresponding Fokker-Planck
equation in the Stratonovich definition, and then we obtain its solutions
for the probability distribution function (PDF).

\textit{Langevin equation and its corresponding Fokker-Planck equation. }We
consider the following Langevin equation in one-dimensional space with a
multiplicative noise term:

\begin{equation}
\dot{\xi}=h(\xi ,t)+g(\xi ,t)\Gamma (t)\text{ ,}  \label{eq1}
\end{equation}%
where $\xi $ is a stochastic variable and $\Gamma (t)$ is the Langevin
force. We assume that the averages $\left\langle \Gamma (t)\right\rangle =0$
and $\left\langle \Gamma (t)\Gamma (\overline{t})\right\rangle =2\delta (t-%
\overline{t})$ \cite{risken}. $h(\xi ,t)$ is the deterministic drift.
Physically, the additive noise ($g(\xi ,t)$ constant) may represent the heat
bath acting on the particle of the system, and the multiplicative noise
term, for variable $g(\xi ,t)$, may represent a fluctuating barrier. For $g=%
\sqrt{D}$ and $h(\xi ,t)=0$, Eq. (\ref{eq1}) describes the Wiener process
and the corresponding probability distribution is described by a Gaussian
function. In the case of $g(\xi ,t)$, some specific functions have been
employed to study, for instance, turbulent flows $(g(x,t)\sim \left\vert
x\right\vert ^{a}t^{b})$ \cite{richar,richar1,hent}. By applying the
Stratonovich approach in a one-dimensional space \cite{risken}, we obtain
the following dynamic equation for the PDF:

\begin{equation}
\frac{\partial W(x,t)}{\partial t}=-\frac{\partial }{\partial x}\left[
D_{1}(x,t)W(x,t)\right] +\frac{\partial ^{2}}{\partial x^{2}}\left[
D_{2}(x,t)W(x,t)\right] \text{ ,}  \label{eq2}
\end{equation}%
where $D_{1}(x,t)$ and $D_{2}(x,t)$ are the drift and diffusion coefficients
given by%
\begin{equation}
D_{1}(x,t)=h(x,t)+\frac{\partial g(x,t)}{\partial x}g(x,t)  \label{eq2a}
\end{equation}%
and%
\begin{equation}
D_{2}(x,t)=g^{2}(x,t)\text{.}  \label{eq2b}
\end{equation}%
We note that Eq. (\ref{eq2}) has \ a spurious drift due to the Stratonovich
definition. Moreover, Eq. (\ref{eq2}) can be written as%
\begin{equation}
\frac{\partial W(x,t)}{\partial t}=-\frac{\partial }{\partial x}\left[
h(x,t)W(x,t)\right] +\frac{\partial }{\partial x}\left[ g(x,t)\frac{\partial
g(x,t)W(x,t)}{\partial x}\right] \text{ .}  \label{eq2c}
\end{equation}%
For the case of $h(x,t)=0$ and $g(x,t)=T(t)D(x)$ the system has been
considered in Ref. \cite{kwok1}; the solution for $W(x,t)$ is given by%
\begin{equation}
W(x,t)=B\left( t\right) \frac{\exp \left[ -\frac{\overline{x}\left( x\right)
^{2}}{4\overline{t}\left( t\right) }\right] }{D(x)\sqrt{\overline{t}\left(
t\right) }}\text{ ,}  \label{eqS1}
\end{equation}%
where%
\begin{equation}
\frac{d\overline{t}}{dt}=T^{2}(t)\text{ ,}  \label{eqS2}
\end{equation}%
\begin{equation}
\frac{d\overline{x}}{dx}=\frac{1}{D(x)}  \label{eqS3}
\end{equation}%
and $B\left( t\right) $ is a normalization factor. Eq. (\ref{eqS1}) can
describe interesting properties such as non-Gaussian distribution,
combination of behaviors like Gaussian (for small distance) and exponential
(for large distance), and combination of behaviors like Gaussian (for small
distance) and power law decay for long distance. Further, it can describe
many bimodal distributions for different forms of $g(x,t)$. For instance,
let us consider $D(x)=\sqrt{D}\left\vert x\right\vert ^{-\frac{\theta }{2}}$%
, then the probability distribution and MSD are given by%
\begin{equation}
W(x,\overline{t})=\left\vert x\right\vert ^{\frac{\theta }{2}}\frac{\exp %
\left[ -\frac{\left\vert x\right\vert ^{2+\theta }}{D(2+\theta )^{2}%
\overline{t}}\right] }{\sqrt{4\pi D\overline{t}}}  \label{eqS4}
\end{equation}%
and%
\begin{equation}
\left\langle x^{2}(t)\right\rangle =\frac{\left[ D^{2}\left( 2+\theta
\right) ^{4}\right] ^{\frac{1}{2+\theta }}\Gamma \left( \frac{6+\theta }{%
2\left( 2+\theta \right) }\right) \overline{t}^{\frac{2}{2+\theta }}\left(
t\right) }{\sqrt{\pi }}\text{, }\theta >-2\text{ .}  \label{eqS5}
\end{equation}%
Moreover, the PDF can also be obtained for $\theta =-2$; in this case the
PDF gives a log-normal distribution. One can see that the multiplicative
noise term in space $D(x)=\sqrt{D}\left\vert x\right\vert ^{-\frac{\theta }{2%
}}$ produces non-Gaussian shapes for the PDF (\ref{eqS4}); it presents a
Gaussian shape only for $\theta =0$. It can also reproduce the asymptotic
behavior of the random-walk model and time fractional dynamic equation for $%
\overline{t}=t^{\beta \left( 2+\theta \right) /2}$ \cite{kwok1}, where $%
0<\beta <1$. Now we want to show two interesting processes which can be
obtained from Eqs. (\ref{eqS4}) and (\ref{eqS5}). To do so, we take $\theta
>-2$. The first one we consider a simple expression for $T(t)$ given by%
\begin{equation}
T(t)=\frac{\sqrt{q}}{\sqrt{t}}\text{, }  \label{eqS6}
\end{equation}%
for $t\gg 1$ . From Eq. (\ref{eqS2}) yields%
\begin{equation}
\overline{t}\left( t\right) =q\ln t\text{ .}  \label{eqS7}
\end{equation}%
Eqs. (\ref{eqS5}) and (\ref{eqS7}) describe the ultraslow diffusion
processes. This kind of diffusion has been found, for instance, in aperiodic
environments \cite{iglo}.

The second one we consider the following $T(t)$:

\begin{equation}
T(t)=\frac{\sqrt{\alpha t^{\alpha -1}}\sqrt{\sum_{j=0}^{n}c_{j}\lambda
_{j}e^{-\lambda _{j}t^{\alpha }}}}{\sum_{i=0}^{n}c_{i}e^{-\lambda
_{i}t^{\alpha }}}\text{ \ ,}  \label{eqS8}
\end{equation}%
where $c_{j}$, $\lambda _{j}$ and $\alpha $ are constants. Using the
function (\ref{eqS8}) one can obtain anomalous diffusion processes with
logarithmic oscillations. We note that the time behavior with logarithmic
oscillation is ubiquitous; examples have been observed, for instance, in
epidemic spreading in fractal media \cite{albano}, financial stock market 
\cite{crash} and diffusion-limited aggregates \cite{sornet}. In Fig. 1 we
show the function $T(t)$ (\ref{eqS8}) for $\lambda _{i}=a^{i}$, $%
c_{i}=(a/b)^{i}$, $a=1/15$ and $b=0.3$; for these values the curves present
logarithmic oscillations with different values of $n$ and $\alpha $. From
Eq. (\ref{eqS2}) we obtain 
\begin{equation}
\overline{t}\left( t\right) =\frac{1}{\sum_{i=0}^{n}c_{i}e^{-\lambda
_{i}t^{\alpha }}}\text{ .}  \label{eqS9}
\end{equation}

Moreover, the PDF (\ref{eqS4}) presents unimodal states for $-2<\theta \leq
0 $ and bimodal states for $\theta >0$ (see in Fig. 2) with pronounced
cusps. The numerical results show that the PDF changes practically nothing
for $n=2$ and $n=6$. In Fig. 3 we show the MSD (\ref{eqS5}) in function of
time $t$; it presents anomalous diffusion processes with logarithmic
oscillations. We see that the main trends in the MSD have power-law
behaviors which indicate subdiffusive regimes.

We now consider that the deterministic drift $h(x,t)$ and multiplicative
noise term $g(x,t)$ are separable in time and space, and they are given by 
\begin{equation}
h(x,t)=T_{1}\left( t\right) D\left( x\right)  \label{eq3}
\end{equation}%
and%
\begin{equation}
g(x,t)=T(t)D(x).  \label{eq4}
\end{equation}%
Then, Eq. (\ref{eq2}) reduces to%
\begin{equation}
\frac{\partial W(x,t)}{\partial t}=-T_{1}\left( t\right) \frac{\partial }{%
\partial x}\left[ D(x)W(x,t)\right] +T^{2}(t)\frac{\partial }{\partial x}%
\left[ D(x)\frac{\partial D(x)W(x,t)}{\partial x}\right] \text{ .}
\label{eq5}
\end{equation}%
We note that the coefficients $h(x,t)$ and $g(x,t)$ given by $%
h(x,t)=g(x,t)=D(x)$ have been used for studying Brownian pumping in
nonequilibrium transport processes \cite{sancho}. By \ suitable
transformations of variables we can show that Eq. (\ref{eq5}) can be reduced
to the constant-diffusion equation without the drift coefficient term. To do
so, we take the following transformations:%
\begin{equation}
\rho \left( x,t\right) =D(x)W(x,t)\text{ ,}  \label{eq6}
\end{equation}%
\begin{equation}
\frac{dt^{\ast }}{dt}=T^{2}(t)  \label{eq7}
\end{equation}%
and%
\begin{equation}
x^{\ast }=\int \frac{\text{d}x}{D(x)}-\int \text{d}tT_{1}\left( t\right) +A%
\text{, }  \label{eq8}
\end{equation}%
where $A$ is a constant, then Eq. (\ref{eq5}) reduces to

\begin{equation}
\frac{\partial \rho \left( t^{\ast },x^{\ast }\right) }{\partial t^{\ast }}=%
\frac{\partial ^{2}\rho \left( t^{\ast },x^{\ast }\right) }{\partial x^{\ast
}{}^{2}}\text{ .}  \label{eq10}
\end{equation}%
Eqs. (\ref{eq7}) and (\ref{eq8}) give the time and space scaling factors
which connect Eq. (\ref{eq5}) to the ordinary diffusion equation (\ref{eq10}%
). Eq. (\ref{eq10}) can be solved and the solution with a natural boundary
condition is given by

\begin{equation}
\rho \left( t^{\ast },x^{\ast }\right) =C\frac{\exp \left[ -\frac{x^{\ast 2}%
}{4t^{\ast }}\right] }{\sqrt{t^{\ast }}}  \label{eq15}
\end{equation}%
where $C$ is a normalization factor. Eqs. (\ref{eq6}) and (\ref{eq15}) show
that the time-dependent coefficients $T(t)$ and $T_{1}(t)$\ do not change
the Gaussian form, however the coefficient $D(x)$ can produce different
forms for the distribution $W(x,t)$ \cite{kwok1}. We note that for $D(x)=%
\sqrt{D}$, $T(t)=1$ and $T_{1}(t)=0$ the Wiener process is recovered.

In order to investigate some details of the solution (\ref{eq15}) we take $%
D(x)=\sqrt{D}\left\vert x\right\vert ^{-\frac{\theta }{2}}$. From Eqs. (\ref%
{eq6}) and (\ref{eq8}), with $A=0$, yields%
\begin{equation*}
W(x,t)=\frac{(-x)^{\frac{\theta }{2}}}{\sqrt{4\pi Dt^{\ast }\left( t\right) }%
}\exp \left[ -\frac{\left( (-x)^{\frac{2+\theta }{2}}+\frac{\sqrt{D}%
(2+\theta )}{2}H(t)\right) ^{2}}{D(2+\theta )^{2}t^{\ast }\left( t\right) }%
\right] ,\text{ \ }x<0\text{ ,}
\end{equation*}%
\begin{equation}
W(x,t)=\frac{x^{\frac{\theta }{2}}}{\sqrt{4\pi Dt^{\ast }\left( t\right) }}%
\exp \left[ -\frac{\left( x^{\frac{2+\theta }{2}}-\frac{\sqrt{D}(2+\theta )}{%
2}H(t)\right) ^{2}}{D(2+\theta )^{2}t^{\ast }\left( t\right) }\right] ,\text{
\ }x>0\text{ ,}  \label{eq22}
\end{equation}%
where $H(t)=\int $d$tT_{1}\left( t\right) $. Eq. (\ref{eq22}) shows that the
drift term produces an asymmetric PDF with respect to the coordinate $x$.
For $T_{1}\left( t\right) =0$ the PDF (\ref{eq22}) reduces to the solution (%
\ref{eqS4}) without the presence of the drift term, and the symmetric PDF is
recovered. In this case, the drift term $T_{1}\left( t\right) $ gives the
duration of this asymmetry. In Fig. 4 we show the asymmetric PDF Eq. (\ref%
{eq22}) for $t=0.2$. The asymmetry of the PDF with $\theta =-0.1$ is more
pronounced than the PDF with $\theta =-0.5$. From Eq. (\ref{eq22}) we obtain%
\begin{equation}
\left\langle x^{2}(t)\right\rangle =\frac{\Gamma \left( \frac{6+\theta }{%
2\left( 2+\theta \right) }\right) }{\sqrt{\pi }}\left[ D\left( 2+\theta
\right) ^{2}t^{\ast }\left( t\right) \right] ^{\frac{2}{2+\theta }}e^{-\frac{%
H^{2}(t)}{t^{\ast }\left( t\right) }}\text{ }_{1}F_{1}\left( \frac{6+\theta 
}{2\left( 2+\theta \right) },\frac{1}{2},\frac{H^{2}(t)}{t^{\ast }\left(
t\right) }\right) \text{,}  \label{eq22a}
\end{equation}%
where $_{1}F_{1}\left( a,b,z\right) $ is the Kummer confluent hypergeometric
function \cite{wolfram}. For $H(t)=0$, without the drift term, we recover
the result Eq. (\ref{eqS5}). Moreover, Eqs. (\ref{eq22}) and (\ref{eq22a})
also present an interesting result; for $H^{2}(t)/t^{\ast }\left( t\right) $
proportional to a constant they give the similar results of Eqs. (\ref{eqS4}%
) and (\ref{eqS5}), without the drift term. In this case, the drift term
only contributes an additional constant to the overall behavior of the
system.

We should note that the solutions (\ref{eq6}) and (\ref{eq15}) can
adequately work for $D(x)$ positive. For $D(x)$ containing negative values
we should modify and take $\rho \left( x,t\right) =-D(x)W(x,t)$ for $D(x)$
negative. For example, let us consider $D(x)=x$. Then we take%
\begin{equation}
\rho \left( x,t\right) =-xW(x,t)\text{, }x<0  \label{eq23}
\end{equation}%
and%
\begin{equation}
\rho \left( x,t\right) =xW(x,t)\text{, }x>0\text{ .}  \label{eq24}
\end{equation}%
From Eq. (\ref{eq8}) we obtain%
\begin{equation}
x^{\ast }=\ln \left\vert x\right\vert -H(t)-\ln \left\vert x_{0}\right\vert
\label{eq24a}
\end{equation}%
and%
\begin{equation}
W(x,t)=\frac{\exp \left[ -\frac{\left( \ln \left\vert x\right\vert -H(t)-\ln
\left\vert x_{0}\right\vert \right) ^{2}}{4t^{\ast }\left( t\right) }\right] 
}{4\sqrt{\pi t^{\ast }\left( t\right) }\left\vert x\right\vert }\text{ .}
\label{eq25}
\end{equation}%
This is the log-normal distribution. The distribution (\ref{eq25}) is the
same as the one given in Ref. \cite{kwok2} for $t^{\ast }\left( t\right) =t$%
, which has been obtained from the method of characteristic. It is worth
mentioning that the coefficients $h(x,t)$ and $g(x,t)$ given by $h(x,t)\sim
x $ and $g(x,t)\sim x$ might be used to investigate the barrier crossing
problem in heavy-ion fusion reaction \cite{mao}, and they can also be used
as a limiting case of a Langevin equation for describing the tumor cell
growth system \cite{ai}.

\textit{Conclusion}. When a multiplicative noise term is introduced into the
simple Langevin equation (\ref{eq1}), even separable in time and space, the
system can exhibit complex behaviors and a rich variety of processes. We
have analytically presented a class of these processes. We hope that they
can be used to mimic a wide class of natural systems.

\bigskip

\textbf{Acknowledgment}

The author acknowledges partial financial support from the Conselho Nacional
de Desenvolvimento Cient\'{\i}fico e Tecnol\'{o}gico (CNPq), Brazilian
agency.

\newpage

\begin{center}
\textbf{FIGURE CAPTIONS}
\end{center}

\bigskip

\bigskip

FIG. 1 - Plots of the function $T(t)$, Eq. (\ref{eqS8}). The dashed lines
correspond to $\alpha =0.5$, whereas the solid lines correspond to $\alpha
=1 $.

\bigskip

FIG. 2 - Plots of the PDF (\ref{eqS4}) for $\lambda _{i}=a^{i}$, $%
c_{i}=(a/b)^{i}$, $a=1/15$, $b=0.3$, $D=1$, $\theta =0.5$ and $\alpha =1$.
The solid lines correspond to $n=2$, whereas the dotted lines correspond to $%
n=6$.

\bigskip

FIG. 3 - Plots of the MSD (\ref{eqS5}) for $\lambda _{i}=a^{i}$, $%
c_{i}=(a/b)^{i}$, $a=1/15$, $b=0.3$, $D=1$ and $\theta =0.5$. The solid line
with $n=4$ corresponds to $\alpha =0.5$, whereas the solid line with $n=6$
corresponds to $\alpha =1$. The dashed lines correspond to the power-law
functions.

\bigskip

FIG. 4 - Plots of the PDF (\ref{eq22}) for $D=1$, $t^{\ast }\left( t\right)
=t$ and $H(t)=t$. The dotted line corresponds to $\theta =-0.5$, whereas the
solid line corresponds to $\theta =-0.1$.

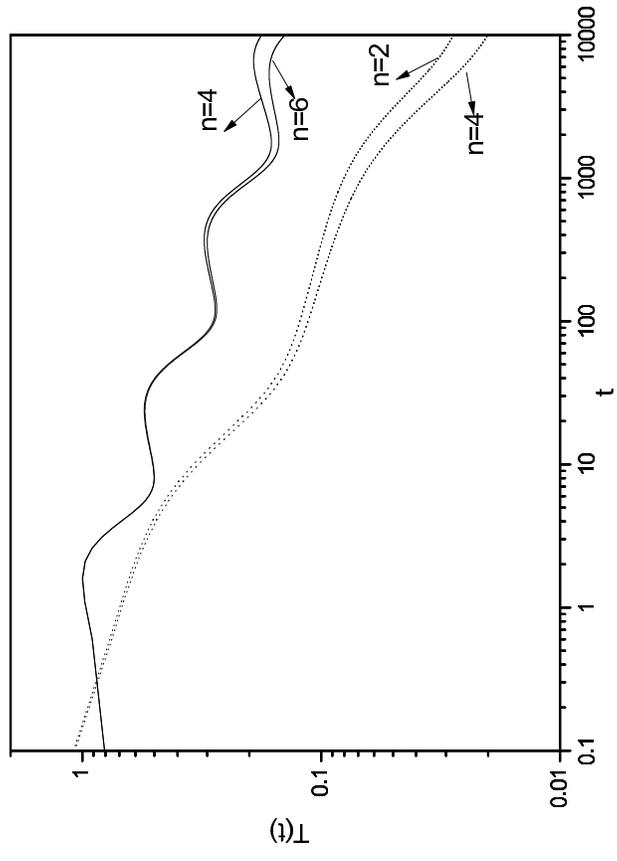
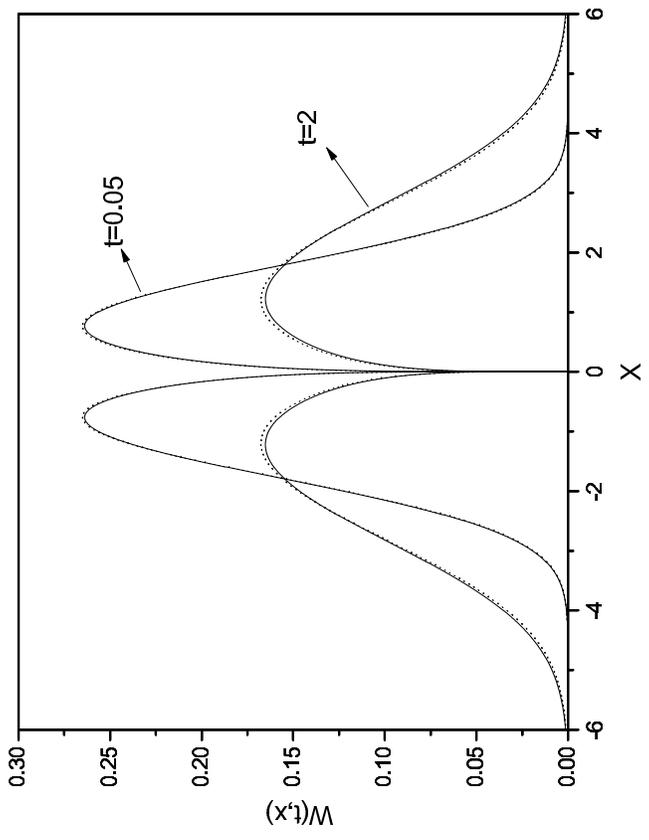
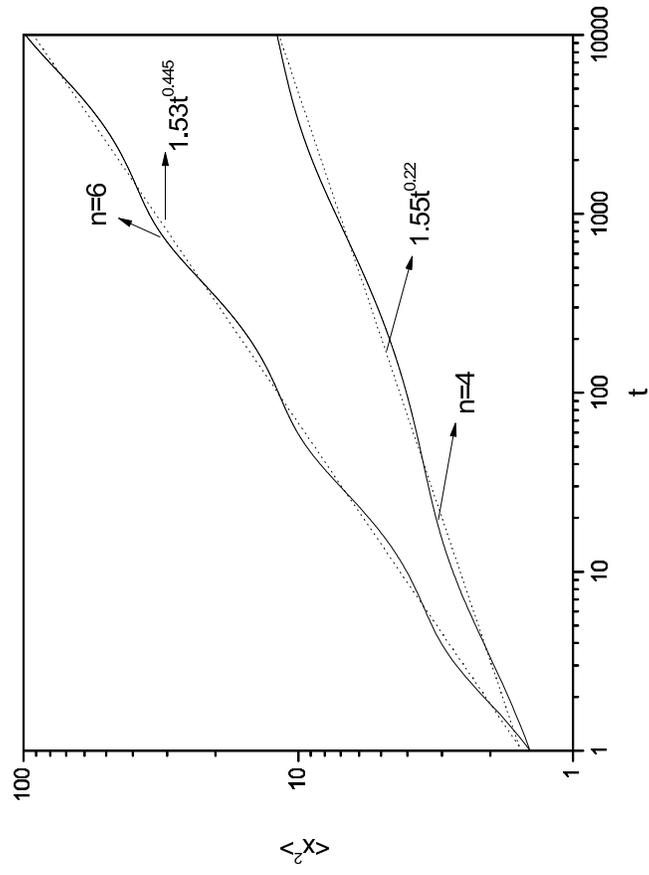
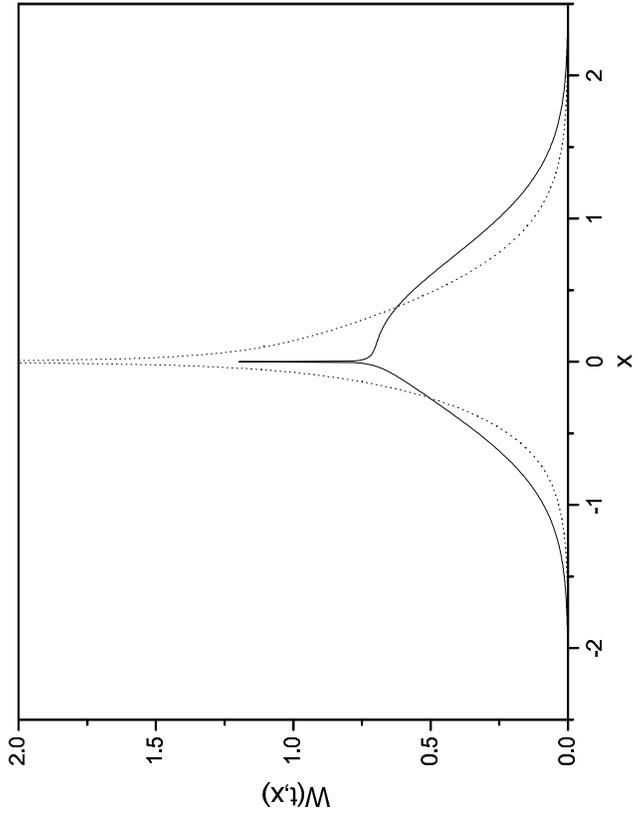
\end{document}